\documentclass{DISproc}

\newcommand{\Compass}{{\sc Compass}}

\newcommand{\Gsi}{{\sc Gsi}}

\newcommand{\Rhic}{{\sc Rhic}}
\newcommand{\Jlab}{{JLab}}

\newcommand{\Jinr}{{\sc Jinr}}
\newcommand{\Jparc}{{\sc Jparc}}

\newcommand{\Hera}{{\sc Hera}}

\newcommand{\dd}{\mathrm{d}}
\newcommand{\RE}{\Re\mathrm{e}}
\newcommand{\IM}{\Im\mathrm{m}}

\begin{document}
\title{Future Programme of COMPASS at CERN}

\author{{\slshape Gerhard K. Mallot}\\[1ex]
on Behalf of the COMPASS Collaboration\\
CERN, 1211 Gen\`eve 23, Switzerland}

\contribID{141}

\doi  

\maketitle

\begin{abstract}
COMPASS at CERN is preparing for a new series of measurements on the nucleon
structure comprising deep virtual Compton scattering and hard exclusive meson
production using muon beams, as well as Drell-Yan reactions using a polarised
proton target and a negative pion beam. The former will mainly constrain the
generalised parton distribution $H$ and determine the transverse size of the
nucleon, while the latter measurements will provide information on
transverse-momentum dependent parton distribution functions. 
The projected results of the programme and the necessary hardware upgrades
are discussed.
\end{abstract}

\section{Introduction}
\label{sec:future}
The \Compass\ Collaboration at CERN proposed in 2010 new measurements on hadron structure
\cite{cp_prop}. The proposal was 
approved in December 2010 and experiments will start in 2012 with a pion/kaon 
polarisability measurement (not discussed here).
The future programme starting 2014 after the accelerator shutdown focuses on 
transverse momentum dependent (TMD) distributions and generalised parton distributions 
(GPDs). A polarised Drell--Yan experiment will take place
in 2014 and in 2015/2016 deeply virtual Compton scattering (DVCS) and hard
exclusive meson production will be studied with a 160\:\GeV\ muon beam and an
unpolarised hydrogen target.
A pilot run for the latter experiment is planned already for late 2012.
In parallel with the GPD programme, high statistics data for semi-inclusive DIS 
will be taken.

\section{GPD programme}
\label{sec:future.GPD}
\begin{wrapfigure}[10]{r}{0.30\textwidth}
  \vspace{-15pt}
  \centering
    \includegraphics[width=0.25\textwidth]{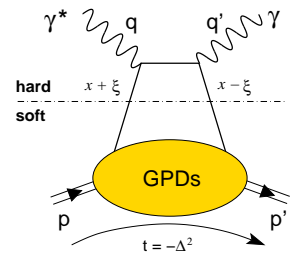}
    \caption{\label{fig:diag_dvcs}
        DVCS process.}
\end{wrapfigure}
The GPDs are universal distributions which contain as limiting cases nucleon form 
factors on the one hand and parton distribution functions (PDFs)
on the other. The GPDs $H^f$ and $\widetilde{H}^f$ ($f=u,d,s,g$) describe 
processes where the nucleon helicity is preserved and contain as limiting cases
the PDFs $f_1$ and $g_1$, respectively. Processes where the nucleon helicity
is flipped are described by the GPDs $E^f$ and $\widetilde{E}^f$ for which 
no such limiting case exists.  GPDs correlate transverse spatial
and longitudinal momentum and thus provide a kind of `nucleon tomography'.
They depend on four variables $x$, $\xi$, $t$, and $Q^2$.
The cleanest process to assess GPDs is DVCS shown in Fig.~\ref{fig:diag_dvcs},
in which also the relevant momentum fractions $x$ (not the Bjorken scaling 
variable) and $\xi$,  and the momentum transfers $t$ and $Q^2$ are defined.
The interest in these distributions was boosted, when X.-D. Ji showed that there
is a sum rule for the total angular momentum $J^f$ of a quark or a gluon 
and the corresponding GPDs
\cite{ji_gpd_sr}.

\begin{figure}[tbp]
  \begin{minipage}[t]{0.47\hsize}
    \vspace{0pt}
    \begin{center}
      \includegraphics[width=\textwidth]{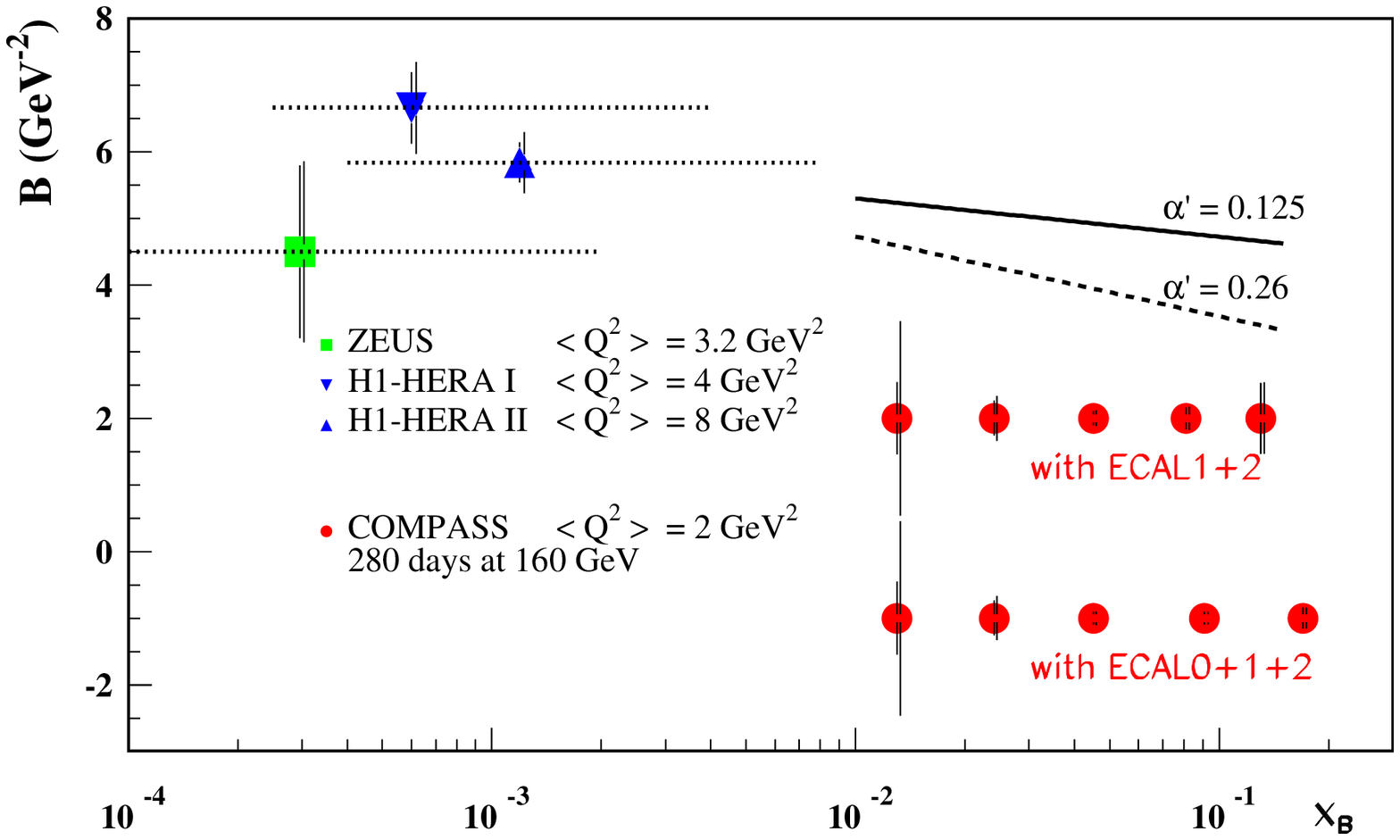}
      \end{center}
    \end{minipage}
  \hfill
  \begin{minipage}[t]{0.47\hsize}
    \vspace{0pt}
    \begin{center}
      \includegraphics[width=\textwidth]{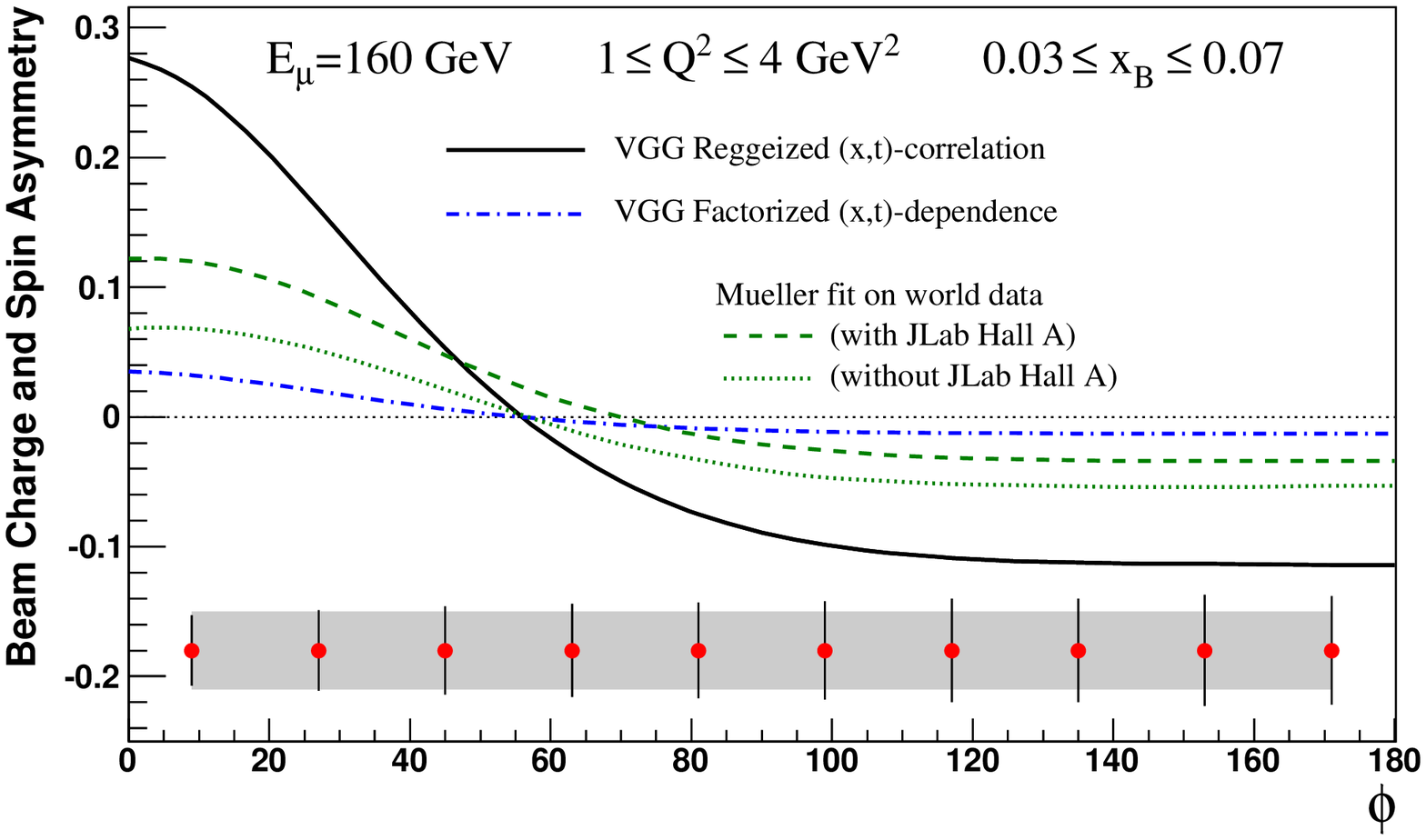}
      \end{center}
    \end{minipage}
  \begin{minipage}[t]{0.47\hsize}
    \vspace{0pt}
      \caption{\label{fig:gpd_t-slope}
        Projected measurements of the $x_B$ dependence of the $t$-slope parameter $B(x_B)$
        (red filled dots)
         using only ECAL1 and ECAL2 (upper row) and with an additional ECAL0 calorimeter (lower row).
         }
    \end{minipage}
  \hfill
  \begin{minipage}[t]{0.47\hsize}
    \vspace{0pt}
      \caption{\label{fig:gpd_BCSA}
        Projected measurements of the  dependence of the
        beam charge-and-spin asymmetry on $\phi$.
        compared to various models from Refs.~\cite{vgg,kmu}}
    \end{minipage}
\end{figure}

The DVCS process interferes with the Bethe--Heitler (BH) process due to identical
final states.
The cross-section then contains five terms 

\begin{equation}
\label{eq:sig_dvcs}
\dd\sigma^{\mu p\rightarrow\mu p \gamma} =
\dd\sigma^\mathrm{BH}
+\dd\sigma_0^\mathrm{DVCS}
+P_\mu \dd\Delta\sigma^\mathrm{DVCS}
+e_\mu \RE\, I
+P_\mu e_\mu\IM\, I,
\end{equation}
where $I$ denotes the DVCS--BH interference term.
An important feature is that the BH contribution can be normalised at small $x_B$, 
where it dominates.
From Eq.~\ref{eq:sig_dvcs} one can build the sum $\cal S$ and difference
$\cal D$ of the $\mu p\rightarrow\mu p \gamma$ cross-section for simultaneous
	change of lepton charge $e_\mu$ and polarisation $P_\mu$ of the incoming lepton beam
($+$ to $-$ and $\leftarrow$ to $\rightarrow$)
\begin{eqnarray}
\label{eq:DS}
{\cal D} &=\dd \sigma^{\stackrel{+}{\leftarrow}} - \dd \sigma^{\stackrel{-}{\rightarrow}}
         & = 2(\dd\sigma_0^\mathrm{DVCS} + \RE\, I)\nonumber\\\
{\cal S} &=\dd \sigma^{\stackrel{+}{\leftarrow}} + \dd \sigma^{\stackrel{-}{\rightarrow}}
         & = 2(\dd\sigma_0^\mathrm{BH} + \dd\sigma_0^\mathrm{DVCS}+\IM\,I).
\end{eqnarray}
The muon beam used at \Compass\ has exactly this behaviour that negative muons have
opposite polarisation than positive muons.
Upon integration over the azimuthal angle $\phi$
the interference contribution to $\cal S$ vanishes \cite{bel2002} and after subtraction 
of the BH contribution one obtains the DVCS cross-section.
This cross-section depends on the squared momentum transfer $t$ from the
initial to final nucleon (Fig.~\ref{fig:diag_dvcs}).
At small $x_B$ one has the relation $\langle r^2_\perp(x_B)\rangle \approx 2B(x_B)$
if the exclusive cross-section is parametrised as $\dd\sigma/\dd t \propto \exp(-B(x_B)|t|$.
The transverse distance $r_\perp$ is measured between the struck quark and the centre of mass
of the spectator system. Thus, independent of any GPD parametrisation, one obtains
a measure of the transverse nucleon size as a function of $x_B$.
Using a parametrisation of the type $B(x_B) = B_0+2\alpha' \log (x_0/x_B)$, one can
characterise the $t$ slope of the cross-section by the parameter $\alpha'$.
The projected precision of a $t$-slope measurement is
presented in Fig.~\ref{fig:gpd_t-slope}. A new electromagnetic calorimeter, ECAL0, will
improve the precision of the measurement and enlarge the accessible range towards larger
$x_B$. Combined with the \Hera\ data and future \Jlab\ data a comprehensive picture of
the evolution of the nucleon's transverse size with $x_B$ will be achieved in a 
model-independent way.
For the 2012 pilot DVCS run we project already a significant
measurement combining the three central $x_B$ bins of Fig.~\ref{fig:gpd_t-slope} into
one large $x_B$.

The $\phi$ dependence of the difference $\cal D$, the sum $\cal S$ and the asymmetry $\cal A=D/S$ 
of the DVCS cross-sections defined in Eq.~\ref{eq:DS} allow for the extraction of quantities 
related to Compton form factors (CFF) which in turn depend on the GPDs. With an unpolarised 
target, \Compass\ DVCS results will mainly provide information on the CFF $\cal H$ and 
thus constrain the GPD $H$. 
Results will be obtained in $(x_B,Q^2)$ bins. An example for
the projected precision in such a bin is shown in Fig.~\ref{fig:gpd_BCSA} for the
beam charge-and-spin asymmetry $\cal A$. 

Some handle on the flavour separation of GPDs may be obtained from hard exclusive
meson production measured simulataneously with DVCS. Here the meson replaces the real photon. 
The GPD $E$ can in principle be assessed using a transversely polarised target. Such measurements
are under consideration for a later stage of the programme. 

Another physics topic pursued in parallel with DVCS, is the study of spin-independent TMD 
distributions like the Boer--Mulders distribution and of fragmentation functions, in particular for strange
quarks. Also the spin-averaged strange quark PDF needs further studies. 

A major rearrangement of the spectrometer target region will be necessary for the
GPD measurements. The polarised target has to be removed and a recoil proton detector,
the Camera detector, will be installed. It consists of two concentric
scintillator barrels of 3.6\:m length and 2.2\:m diameter for the outer barrel.
The photomultiplier signals will be digitised with 1\:GHz to cope with the high rate and 
pile-up. Camera is essential to ensure the exclusivity of the observed reactions. It houses
on the central axis a 2.5\:m long liquid hydrogen target.
In order to improve the acceptance of real photons, a third electromagnetic calorimeter,
ECAL0, will be constructed and placed just downstream of the Camera detector.
Multipixel avalanche photodiodes were chosen for the readout to avoid problems due the
magnetic field of the close spectrometer magnet SM1. Furthermore the RICH photodetection
will be improved.

\section{Drell--Yan programme}
\label{sec:future.DY}
\begin{wrapfigure}{r}{0.45\textwidth}
  \centering
  \includegraphics[width=0.45\textwidth]{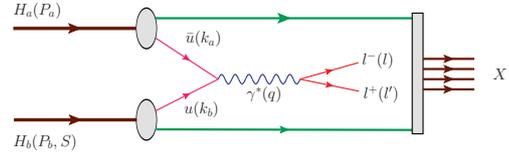}
  \caption{\label{fig:diag_dy}
    Sketch of the Drell--Yan process.
    }
\end{wrapfigure}
The second approach to access transverse nucleon structure in the future \Compass\ programme
is via the Drell--Yan process (Fig.~\ref{fig:diag_dy}) using a 190\:\GeV\ negative pion beam 
impinging on a transversely polarised proton target (NH$_3$). 
The advantage of DY processes
is that fragmentation functions are not involved. However, this has to be paid by a convolution
of two distribution functions. 
The DY cross-section is given by
$\sigma^\mathrm{DY}\propto \sum_f f_{\bar u|\pi^-}\otimes f'_{u|p}$, where $f$ and $f'$ are
generic place holders of PDFs. For $\pi^-p$ scattering the process is dominated by the up quark distributions. 
Polarised DY experiments can study TMD distributions like the Sivers and Boer--Mulders distributions.
Theory predicts that these naive $T$-odd TMD distributions obey a restricted universality
and change sign when observed in SIDIS and DY
\begin{equation}
\left.f_{1T}^\perp\right|_{DY}=\left.-f_{1T}^\perp\right|_{DIS}
\hskip 1cm \mbox{and}\hskip 1cm 
\left.h_{1}^\perp\right|_{DY}=\left.-h_{1}^\perp\right|_{DIS}.
\end{equation}
This sign change is due to switching from final-state interaction in SIDIS to
initial-state interaction in DY
\cite{SIDIS-DY_collins}. 
A violation of this prediction would imply drastic consequences on how cross-sections are calculated.
This has generated wide-spread interest in a  direct comparison of TMD distributions obtained from 
SIDIS and DY, respectively. Plans for future polarised DY experiments exist at various laboratories, e.g.\ 
at \Rhic, \Jparc, \Gsi\ and at \Jinr.
The \Compass\ DY experiment is planned and approved for 2014 and
primarily
assess transversity $h_1$ and the $T$-odd Sivers and Boer--Mulders TMD distributions, $f_{1T}^\perp$ and 
$h_1^\perp$ for up quarks. For all of these \Compass\ SIDIS measurements exist, showing non-vanishing
asymmetries for the proton.
 
\begin{wrapfigure}[24]{r}{0.50\textwidth}
  \vspace{-10pt}
  \centering
  \includegraphics[width=0.50\textwidth]{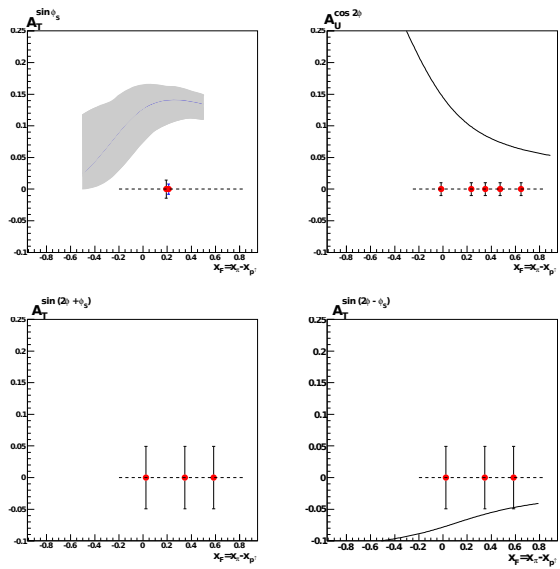}
  \caption{\label{fig:dy_asym}
        Projections for the asymmetries $A_T^{\sin\phi_S}$ (Sivers), $A_U^{2\cos\phi}$ (Boer--Mulders), 
        $A_T^{\sin(2\phi+\phi_S)}$ and
        $A_T^{\sin(2\phi-\phi_S)}$.
        }
\end{wrapfigure}
To avoid the $J/\psi$ region and the region of background from charm decays, the experiments will focus on
dimuon masses $4\:\GeV<M_{\mu\mu}<9\:\GeV$. 
The azimuthal asymmetries depend on two azimuthal angles, $\phi_S$ of the target spin with respect to the
transverse momentum of the virtual photon in the target rest frame and $\phi$ between the incoming hadron 
and outgoing lepton plane in the Collins--Soper frame, a the polar angle $\theta$ of
the lepton pair (see Ref.~\cite{cp_prop}), as well as 
on the Feynman variable $x_F = x_\pi-x_p$. Here $x_\pi$ and $x_p$ 
are the momentum fractions carried by the involved quarks in the pion and proton, respectively.
The projected $A_T^{\sin\phi_S}$ asymmetry measurement in the high-mass region $4\:\GeV<M_{\mu\mu}<9\:\GeV$ is 
compared to predictions
in Fig.~\ref{fig:dy_asym} (top left). The measurement will certainly be able to answer
the sign question of $T$-odd TMD distributions and allow for a comparison of the absolute size
of the effects in SIDIS and DY. However, a determination of the shape of the Sivers TMD distribution
in DY will only be possible with further measurements. The shaded grey area and the central line 
in Fig.~\ref{fig:dy_asym}
correspond to a calculation based on a TMD PDF fit to data \cite{dy_theo_Ans}. 
The Boer--Mulders related asymmetry $A_T\cos2\phi$ will
be determined with high precision.

As the measurement is statistics limited, optimising luminosity is mandatory.
A massive hadron absorber downstream of the target reduces radiation and detector occupancy
problems.
Therefore the polarised target has to
be moved upstream
by about 3\:m.
The absorber consists of a tungsten
core surrounded by alumina ($\mathrm{Al_2O_3}$), which minimises
multiple scattering. This is essential to disentangle the oppositely polarised
target cells in the track reconstruction.

{\raggedright
\begin{footnotesize}

\end{footnotesize}
}

\end{document}